\documentclass{amsproc}
\usepackage{amssymb}

\usepackage{graphicx}
\usepackage{amsmath}

\input{tcilatex}

\begin{document}

\title{\textbf{TOPOLOGICAL\ QUANTUM\ FIELD\ THEORIES}}
\author{Albert Schwarz}

\address{Department of Mathematics,
University of California at Davis,\\
Davis, CA 95616}

\email{schwarz@math.ucdavis.edu}

\subjclass{81T}

\keywords{TQFT, Chern-Simons, Batalin-Vilkovisky}

\begin{abstract}

Following my plenary lecture on ICMP2000 I  review  my results
concerning two closely
related topics: topological quantum field theories and the problem of
quantization of gauge theories. I start with old results (first examples
of topological quantum field theories were constructed in my papers in late
seventies) and I come to some new results, that were not published yet.
\end{abstract}

\maketitle

\bigskip

\textbf{0. Introduction.}

I review  my results concerning two closely
related topics: topological quantum field theories and the problem of
quantization of gauge theories. I'll start with old results (first examples
of topological quantum field theories were constructed in my papers in late
seventies) and I'll come to some new results, that were not published yet.
But first of all I would like to give a short (and very incomplete) overview
of these problems and of some related questions (see, for example ,
\cite{BBRT}, \cite{LL} for more
complete review).

Massive intervention of topology into quantum field theory was triggered by
discovery of magnetic monopoles in $SU(2)$ gauge theory with scalar
fields-Georgi-Glashow model (Polyakov \cite{P}, 'tHooft \cite{H}). It was
recognized very soon (\cite{57}, \cite{65}, \cite{69}, \cite{MP},
\cite{AF})
that magnetic charge has topological nature and that simple topological
considerations can be used to prove the existence of magnetic monopoles in a
large class of gauge theories (including all grand unification theories ).

Other topologically non-trivial field configurations were considered shortly
.The most significant role was played by topologically non-trivial extremals
of Yang-Mills Euclidean action-gauge instantons \cite{64}.

I started as a topologist in fifties, and it was very pleasant for me to
find important applications of topological ideas to physics-homology and
homotopy theory, characteristic classes, Atiyah-Singer index theory became
common tools in quantum field theory. I was pleased even more when I found
an idea permitting to apply quantum field theory to topology. The idea was
very simple-if an action functional depends only on smooth structure of a
manifold then corresponding physical quantities (in particular, the
partition function) should have the same property. The simplest example is a
functional 
\begin{equation}
S=\int_{M}A\wedge dA,  \tag{1}
\end{equation}%
where $A$ is a $1$-form on three-dimensional compact manifold $M$. This
functional is invariant with respect to gauge transformations $A\rightarrow
A+d\lambda ,$ therefore to calculate its partition function one should
impose gauge condition. The gauge condition cannot be invariant with respect
to diffeomorphisms; it should involve some additional data, for example
Riemannian metric. However, the answer should not depend on the choice of
gauge condition. This logic is not flawless; formal arguments above can be
destroyed by quantum anomalies. However, it is possible to give a rigorous
theory of partition function of degenerate quadratic functionals and to
relate the partition function of the functional (1) and of its
multidimensional generalizations to Ray-Singer torsion-smooth version of
Reidemeister torsion.(See \cite{80}, \cite{77} and Sec.2 )

The theory\ described by the action functional (1) is the simplest example
of topological quantum field theory. It admits a non-abelian
generalization-so called Chern-Simons action functional 
\begin{equation}
S=\int_{M}\frac{1}{2}\func{Tr}A\wedge dA+\frac{1}{3}\func{Tr}A\wedge A\wedge
A  \tag{2}
\end{equation}%
where $A$ is a matrix-valued $1$-form on three-dimensional compact manifold $%
M$. (The functional (1) is called sometimes abelian Chern-Simons action.) It
was conjectured in \cite{BAKU} that the action functional (2) leads to
invariants of manifold $M$ that are closely related to Jones polynomial of
knots. I was not able to prove this conjecture. This was done a year later
in remarkable paper by E.\ Witten\ \cite{W4}, who found a way to calculate
physical quantities associated with Chern-Simons action in terms of
two-dimensional conformal field theory. It is difficult to overestimate the
influence of this paper and of the papers \cite{W2}, \cite{W3} that E.
Witten
has written almost at the same time. It became clear after these papers that
using topological quantum field theories one can obtain very interesting
mathematical results and that these theories are very useful not only in
mathematics, but also in physics. The papers \cite{W2}, \cite{W3} are
closely
related to earlier Witten's\ paper \cite{W1} and to work of Donaldson, Floer
and Gromov \cite{D}, \cite{F}, \cite{F2}, \cite{G}. Donaldson and Floer
used
some ideas from physics to obtain beautiful mathematical results. Donaldson
applied instanton moduli space, studied earlier in \cite{73}, \cite{82}, 
\cite{ahs} to obtain very strong results about 4-manifolds. Floer's starting
point was Witten's paper about supersymmetry and Morse theory \cite{W1}.
Witten has shown that their constructions, as well as Gromov's invariants of
symplectic manifolds, can be understood in the framework of topological
quantum field theory. This understanding led to very important development
in pure mathematics culminating in theory of Seiberg-Witten invariants of
4-manifolds based on results of \cite{sw}, \cite{W5}and in enumerative
geometry of (pseudo)holomorphic curves on symplectic manifolds \cite{can},%
\cite{kon}, \cite{giv}. After axiomatization of topological quantum field
theories by Atiyah \cite{A2}these theories became a subject of extensive
mathematical analysis; I'll stay mostly on Lagrangian viewpoint in this
lecture.

I would like to mention a couple of important ideas that appeared in
Witten's papers. One of them is to allow dependence of action functional $S$
of metric $g_{\alpha \beta },$ but to require the energy- momentum tensor $%
T^{\alpha \beta }=\delta S/\delta g_{\alpha \beta }$ that governs the
dependence of $S$ of the metric to be BRST-trivial (See Sec.3 for
introduction to BRST formalism.) Then expectation values of observables (of
BRST closed functionals) should be metric independent, i.e. the quantum
field theory is topological. One calls theories of this kind topological
quantum theories of Witten type, as opposed to topological theories of
Schwarz type, when the action functional is metric independent. The next
idea that in supersymmetric theories one can declare a supersymmetry
generator that squares to zero to be a BRST operator. Then it follows from
supersymmetry algebra that the translation operator is BRST-trivial; this
means that correlation functions $<A_{1}(x_{1})...A_{n}(x_{n})>$ of BRST
closed observables $A_{1},...,A_{n}$ don't depend on $x_{1},...,x_{n}$. 
One
can say that BRST closed observables are topological observables: it is
possible construct a topological quantum field theory where correlation
functions are equal to correlation functions of topological observables of
sypersymmetric theory. One says that topological theory is obtained from
supersymmetric theory by means of twisting. Twisting $N=2$ supersymmetric
four-dimensional gauge theory one can obtain topological theory that is
closely related to Donaldson invariants of four-dimensional manifolds.
Two-dimensional sigma-model has $N=2$ superconformal symmetry if the target
space is a Kaehler manifold. There are two essentially different
possibilities to choose BRST-operator and to twist the theory. (This is true
for any $N=2$ superconformal theory; see \cite{6} for discussion in the
framework of axiomatic quantum field theory.) In one case we obtain so
called $A$-model; it is defined for every symplectic target and its
correlation function can be expressed in terms of (pseudo)holomorphic curves
(in terms of Gromov-Witten invariants). In an other case we obtain $B$-model
where one should assume that the target is a complex Calabi-Yau manifold.
We'll not discuss remarkable theory relating $A$-model on one manifold with $%
B$-theory on another (mirror) manifold (\cite{can}, \cite{kon}, 
\cite{giv}).

\bigskip

As I emphasized from the very beginning the development of topological
quantum field theory was intertwined with the progress in our understanding
of quantization of gauge theories. To analyze topological theories related
to Ray-Singer torsion it was necessary to deal with so called reducible
gauge theories; the analysis of this class of gauge theories was given in %
\cite{77}. In this case the needs of TQFT led to progress in the problem of
quantization. (See Sec.1 and 2) Later, as we have seen, BRST formalism in
quantization of gauge theories was used to construct topological quantum
theories of new type. It seems that Batalin-Vilkovisky version of BRST
formalism is very convenient to construct action functionals of TQFT. I'll
give a short exposition of Batalin-Vilkovisky formalism from geometric
viewpoint following my papers \cite{2}, \cite{4} \cite{7}. This geometric
approach will be used to construct BV topological sigma-model, that includes
many interesting topological quantum field theories as particular cases \cite%
{aksz}. In particular, it includes BV version of Chern-Simons action
functional as well as its multidimensional generalization (Sec.4). It is
interesting to notice that some of Witten type topological theories can be
formulated as theories with metric independent action in BV formalism; may
be this is true for all theories of this kind. Sec.5 devoted to perturbation
theory in BV-formalism. It seems that the version of perturbation expansion
described in this section was not analyzed in previous publications
although
it was used implicitly in [8]. \ In Sec.6 I'll discuss recent results about
quantum observables in BV formalism and their application to TQFT following
\ \cite{999}.

In Sec.7 I'll formulate some new results about families of action
functionals. It will be shown that using families of equivalent action
functionals or families of gauge conditions one can construct some numbers
generalizing expectation values of observables. More precisely, one can
consider a kind of moduli space corresponding to a given action functional
and under certain conditions one can define closed differential forms on
moduli space. Integrating these forms over cycles we obtain some interesting
quantities. In particular, one can show that the application of this
construction to two-dimensional topological quantum field theory gives
string amplitudes. Applying it to multi-dimensional analog \ of Chern-Simons
action functional we obtain cohomology classes of \ BDiff $(M)$ constructed
by M.Kontsevich.

The results I know in this direction are not complete. Some of them were
obtained in collaboration with M. Kontsevich. Several years ago we planned
to work together on families of topological quantum field theories,
however
both of us found more urgent problems to study.

\bigskip

\textbf{1. Quantization of gauge theories.}

\textbf{\bigskip }

Let us consider a functional $S$\ defined on space $\mathcal{E}$ \ (''space
of fields''). We can regard $S$\ as classical action functional;
Euler-Lagrange equations for stationary points of $S$\ are interpreted as
equations of motion of corresponding classical mechanical system. We can
also try to consider corresponding \ quantum system; this means that we
should calculate the integral \ $Z$ of \ $\exp (-S)$ over \ $\mathcal{E}$\
(the partition function) and the expressions of the form $Z_{A}=\int_{%
\mathcal{E}}A\exp (-S)$\ \ where $A$\ is a function on $\mathcal{E}$\ .\ \
(One can regard $Z_{A}/Z$\ as an expectation value of the observable $A$ ).
Notice, that in the above terminology we consider $A$ as a Euclidean action
functional.

\bigskip

In interesting cases the space $\mathcal{E}$ is infinite-dimensional,
therefore the integrals $Z$ and $Z_{A}$ are ill-defined; it is quite
difficult to make sense of them. However, if $\mathcal{E}$ is a vector space
and $S$ is represented as a sum of quadratic \ functional $S_{qu}$\ and
polynomial \ functional $V$\ one can try to construct perturbation series
for these integrals, considering $V$ as a perturbation. This problem is much
easier, but still it is not simple. (To solve it one should analyze the
integrals of $AV^{n}\exp (-S_{qu})$ over $\mathcal{E}$.) If the functional $%
S $ has a large symmetry group (gauge group), the quadratic part of $S$ is
degenerate and in addition to other problems we should deal with the
divergence of gaussian integral $\int_{\mathcal{E}}\exp (-S_{qu})$.

\bigskip

The standard way to work with ill-defined infinite-dimensional\ \ integrals
is to consider similar finite-dimensional\ \ integrals and to apply
rigorously proven finite-dimensional\ formulas to \ infinite-dimensional\
case without any justification. (Such a justification is impossible because
usually we don't have any rigorous definitions of \ integrals at hand.
Sometimes it is useful to say that finite-dimensional\ theorems become
definitions in \ infinite-dimensional\ \ case ).

\bigskip

Let us consider a compact Lie group $G$ that acts on finite-dimensional\
Riemannian manifold $M$\ preserving the Riemannian metric. Let us fix a $G$%
-invariant function $f$ on $M$. If $G$ acts freely on $M$ and $\Sigma $ is
a
subset of $M$ having precisely one common point with every orbit of $G$,
then we can reduce the integration over $M$ to the integration over $\Sigma
. $ (In physics $G$\ plays the role of gauge group; replacing $M$\ by $%
\Sigma $\ we impose gauge condition.)\ \ \bigskip More precisely, if the set 
$\Sigma $ is singled out by means of equation $F(x)=0$ where $%
F==(F^{1},...,F^{k})$ is a map of $M$ into $\mathbb{R}^{k}$, then

\bigskip 
\begin{equation*}
\int_{M}\exp (-S)d\mu =\int_{M}\exp (-S)W_{F}(x)\delta (F(x))d\mu
\end{equation*}
where $W_{F}(x)$ is specified by the formula

\begin{equation}
W_{F}(x)\cdot \int \delta (F(gx))dg=1.  \tag{4}
\end{equation}
(Here $dg$ stands for invariant volume element on $G$ normalized by the
condition that the volume of $G$ is equal to $1$.)

If $M$ is an infinite-dimensional manifold one can consider the right hand
side of (3) as a definition of left hand side. This idea (Faddeev-Popov
trick) is very useful in quantization of gauge theories. Of course, to apply
this idea one should verify that the left hand side does not depend on the
choice of $\Sigma $ (on the choice of gauge condition).

\bigskip

The physical quantities are defined as functional integrals of $A\exp (-S)$
where both $S$ and $A$ are gauge invariant. In many cases Faddeev-Popov
trick
permits us to obtain a perturbation series for quantities of this kind.

\bigskip

We will use an invariant form of Faddev-Popov trick that is based on the
following finite-dimensional statement.

Let us consider again a compact group $G$ of isometries of Riemannian
manifold $M$ and a $G$-invariant function $f$. Without loss of generality we
can assume that stable subgroups $H_{x}$ for all points $x\in M$ are
conjugate (this fact is always true for almost all points). Then

\begin{equation}
\int_{M}f(x)d\mu =\int_{M/G}f(x)(D(x))^{1/2}(V(H_{x}))^{-1}d\nu  \tag{5}
\end{equation}

Here $d\nu $ is the volume element corresponding to the natural Riemannian
metric on $M/G,V(H_{x})$ stands for the volume of $H_{x}$ in the metric
induced by invariant Riemannian metric on $G$ obeying $V(G)=1$ and $%
D(x)=\det \widetilde{\mathcal{T}_{x}^{+}}\widetilde{\mathcal{T}_{x}}$ where $%
\widetilde{\mathcal{T}_{x}}$ is a linear operator acting from Lie $G/$Lie $%
H_{x}$ into tangent space $T_{x}(M)$. (The action of $G$ on $M$ determines
an operator $\mathcal{T}_{x}$: Lie $G\rightarrow T_{x}(M)$. This operator
descends to $\widetilde{T_{x}}:$ Lie $G/$Lie $H_{x}\rightarrow T_{x}(M)$.)
Let us say that the compact Lie groups $G_{0},...,G_{N}$ and homomorphisms $%
T_{i}:G_{i}\rightarrow G_{i-1}$ form a resolution of subgroup $H\subset G$
if $G=G_{0},\func{Im}T_{1}=H,\func{Im}T_{i+1}=KerT_{i}$. We introduce an
invariant Riemannian metric on $G_{i}$, and assume that it is normalized by
the condition $V(G_{i})=1$. The homomorphism $T_{i}$ generates a
homomorphism $\mathcal{T}_{i}$ of corresponding Lie algebras; it descends to
a linear map $\widetilde{\mathcal{T}_{i}}:$Lie $G_{i}/Ker\mathcal{T}%
_{i}\rightarrow $Lie $G_{i-1}$. Using (5) it is easy to check that

\begin{equation}
\log V(H)=\sum \frac{1}{2}(-1)^{i-1}\log \det \mathcal{T}_{i}^{\ast }%
\mathcal{T}_{i}.  \tag{6}
\end{equation}

Combining (5),(6) we obtain 
\begin{eqnarray}
\int_{M}f(x)d\mu &=&\int_{M/G}f(x)(\det \widetilde{\mathcal{T}_{x}^{\ast }}%
\widetilde{\mathcal{T}_{x}})^{1/2}\Pi _{1\leq i\leq N}(\det \widetilde{%
\mathcal{T}_{i}^{\ast }}\widetilde{\mathcal{T}_{i}})^{\frac{1}{2}%
(-1)^{i}}d\nu =  \TCItag{7} \\
&&\int_{M/G}f(x)(\det \square _{0}(x))^{1/2}\Pi _{1\leq i\leq N}(\det
\square _{i})^{\sigma _{i}}d\nu  \notag
\end{eqnarray}

$\bigskip $where $\square _{0}(x)=\mathcal{T}_{x}^{\ast }\mathcal{T}_{x}+%
\mathcal{T}_{1}\mathcal{T}_{1}^{\ast }$, $\square _{i}=\mathcal{T}_{i}^{\ast
}\mathcal{T}_{i}\mathcal{+T}_{i+1}\mathcal{T}_{i+1}^{\ast }$, $\sigma _{i}=%
\frac{1}{2}(-1)^{i}/(i+1)$.

\bigskip

Let us consider the case when the function $f(x)$ is defined on vector space 
$\mathcal{E}$ and has the form $f((x)=\exp (-\mathcal{S}(x))$ where $%
\mathcal{S}(x)$ is a quadratic functional. If $\mathcal{E}$ is equipped with
inner product we can represent $\mathcal{S}(x)$ in the form

\begin{equation*}
\mathcal{S}(x)=<Sx,x>=<x,Sx>.
\end{equation*}

\bigskip In finite-dimensional case

\bigskip 
\begin{equation}
\int_{\mathcal{E}}e^{-<Sx,x>}dx=(\det S)^{-1/2}  \tag{8}
\end{equation}
for appropriate normalization of the volume element on $\mathcal{E}$. If $S$
is non-degenerate we accept the right hand side of (8)\ as a definition of
infinite-dimensional Gaussian integral. However, to apply this definition we
should have a definition of infinite-dimensional determinant. One of
possible approaches is based on the notion of zeta-function:

\begin{equation*}
\log \det S=\frac{1}{2}\log \det S^{\ast }S=-\frac{1}{2}\varsigma _{S^{\ast
}S}^{\prime }(0).
\end{equation*}

(One can define the zeta-function of nonnegative operator $A$ by the formula 
$\varsigma _{A}(s)=\sum \lambda _{i}^{-s}$, where $\lambda _{i}$ runs over
positive eigenvalues of $A$. In the case when $A$ is an elliptic operator on
compact manifold the series converges for $s\gg 0$, however one can define $%
\varsigma _{A}^{\prime }(0)=d\varsigma _{A}(s)/ds|_{s=0}$ by means of
analytic continuation. Notice that this definition can be applied also to
operators having zero modes, because zero eigenvalues don't enter the
expression for zeta-function.)

\bigskip

If the functional $\mathcal{S}$ is degenerate, one can try to define the
Gaussian functional integral (the partition function corresponding to $%
\mathcal{S}$) by the formula (8). However, such an attempt does not lead to
interesting results; we need additional structure to give a reasonable
definition.

\bigskip

Let us consider a quadratic functional $\mathcal{S}$ on the space $\mathcal{%
E=E}_{0}$, vector spaces $\mathcal{E}_{1},...,\mathcal{E}_{n}$and operators $%
T_{i}:\mathcal{E}_{i}\rightarrow \mathcal{E}_{i-1}$, obeying $T_{i-1}\cdot
T_{i}=0,$ $\mathcal{S}(x+T_{1}y)=\mathcal{S}(x)$. We will assume that spaces 
$\mathcal{E}_{i}$ are equipped with Hermitian inner product; this means that
we can represent $\mathcal{S}(x)$ in the form $\mathcal{S}(x)=<Sx,x>=<x,Sx>$
and consider adjoint operators $T_{i}^{\ast }.$

\bigskip

We will say that the spaces $\mathcal{E}_{i}$ and operators $\mathcal{T}_{i}$
constitute on ellliptic resolution of the functional $\mathcal{S}$ if the
space $\mathcal{E}_{i},$ $i=0,...,N,$ can be considered as a space smooth
sections of vector bundle with compact base and the operators $\square
_{0}=S^{2}+T_{1}T_{1}^{\ast },$ $\square _{i}=T_{i}^{\ast
}T_{i}+T_{i+1}T_{i+1}^{\ast }$ are elliptic operators. We define the
partition function of the functional $\mathcal{S}$ with respect to elliptic
resolution by the formula:

\begin{equation}
Z=\Pi _{0\leq i\leq N}(\det \square _{i})^{(-1)^{i+1}(2i+1)/4}.  \tag{9}
\end{equation}

\bigskip

(We can come to this definition applying formally Eqn (7)). In general, $Z\; 
$depends on the choice of inner products on $\mathcal{E}_{0},...,\mathcal{E}%
_{N}$. However, it is possible to calculate the variation of $Z$ when
these
inner products$\;$vary. Let us suppose that we have a family $<,>_{i}^{u}$of
inner products on $\mathcal{E}_{i}\;$depending om parameter $u\;$denote the
operators governing infinitesimal variation$\;$of inner product by $%
B_{i}^{u}:\;\;\;$

\begin{equation*}
\frac{d}{du}<f,g>_{i}^{u}=<B_{i}^{u}f,g>_{i}^{u}=<f,B_{i}^{u}g>_{i}^{u}.
\end{equation*}

Then

\begin{equation*}
\frac{d\log Z(u)}{du}=\frac{1}{2}\sum_{0\leq i\leq n}(-1)^{i}\Psi
_{0}(B_{i}^{u}|\square _{i}^{u}).
\end{equation*}
\newline

\bigskip (The Seeley coefficients $\Psi _{k}(R|A)$ are defined by means of
asymptotic expansion:

\begin{equation*}
Tr(R\text{ }e^{-At})=\sum \Psi _{k}(R|A)t^{-k}
\end{equation*}
for $t\rightarrow 0$). We assume that the operators $\square _{i}$ do not
have zero modes (the resolution is an exact sequence); to take zero modes
into account one should subtract the trace of $B_{i}$ on the space of zero
modes from $\Psi _{0}$.

If $\square _{i}$ are differrential operators then the Seeley coefficients
are given by local formulas. If the operators $\square _{i}$ act on vector
bundle with odd dimensional base then $\Psi _{0}$ vanishes and $Z$ does not
depend on $u$. In other words, we don't have quantum anomaly in this case.
(One speaks about quantum anomaly if something that is true for classical
theory is violated at quantum level. Classical theory is determined by the
functional $\mathcal{S}$ in our case; it does not depend on inner products.
Therefore dependence of inner products can be characterized as quantum
anomaly.)

Notice that in this statement we assumed that the operators $\square _{i}$
don't have zero modes. In general we have to consider a vector space $%
\mathcal{H}=\sum \mathcal{H}_{i}=\sum KerT_{i}/ImT_{i+1}\approx \sum
Ker\Delta _{i}$ (homology of the resolution $(\mathcal{E}_{i},T_{i})$). The
partition function can be regarded as a measure on linear superspace $%
\mathcal{H}$ (natural $\mathbb{Z}_{2}$-grading determines a structure of
superspace on $\mathcal{H}$). If the relevant Seeley coefficients vanish the
partition function does not depend on the choice of inner products on $%
\mathcal{E}_{i}$ (see\cite{77} \cite{109}).

\bigskip

\textbf{2. Topological gauge theories.}

\bigskip

Let us consider an action functional

\begin{equation}
\mathcal{S}=\int_{M}A\wedge dA  \tag{10}
\end{equation}
where $M$ is a compact $(2n+1)$-dimensional manifold, $A$ is an $n$-form on $%
M$. This functional is invariant with respect to transformations $%
A\rightarrow A+d\lambda $, where $\lambda $ is an $(n-1)$-form on $M$.
Denoting the space of smooth $k$-forms on $M$ by $\Omega ^{k}$ we can say
that $\mathcal{S}$ is a $\deg $enerate quadratic functional on $\Omega ^{n}$
and that $\Omega ^{n-1}$ can be considerate as symmetry group of (10). In
the case $n=1$ we can use the Faddeev-Popov trick to calculate the
corresponding partition function. If $n>1$ then the map $d:\Omega
^{n-1}\rightarrow \Omega ^{n}$ has an infinite-dimensional kernel; in this
case we can use the notion of elliptic resolution to define the partition
function of (10). Namely, the spaces $\Omega ^{n},\Omega ^{n-1},...,\Omega
^{0}$ and operators $d:\Omega ^{k-1}\rightarrow \Omega ^{k}$ provide us with
elliptic resolution of (10).

Let us fix a Riemannian metric on $M$. This metric induces an inner product
on spaces $\Omega ^{k}$; we can use this inner product to calculate the
partition function. Using (9) we obtain an expression of the partition
function $Z$ in terms of det$\Delta _{k},$where $\Delta _{k}=d^{\ast
}d+dd^{\ast }$ stands for the Laplace operator on the space of $k$-forms.

The same construction can be repeated in the case when we allow forms with
coefficients in a local system (i.e. forms taking values in fibres of a
vector bundle equipped with a flat connection).

In the acyclic case (in the case when the operators $\Delta _{k}$ have no
zero modes) it follows from the results of Sec.1 that $Z$ does not depends
on Riemannian metric on $M$. This means that $Z$ is invariant with respect
to diffeomorphisms. It is easy to check that $Z$ coincides with Ray-Singer
torsion (smooth version of Reidemeister torsion) and therefore is a
topological invariant. In general case $Z$ is a topologically invariant
measure on superspace $H(M)$ (direct sum of cohomology groups of $M$).

\bigskip

In the case $\dim M=3$ one can generalize the action functional (10) to the
case when $A$ is a $1$-form taking values in Lie algebra $\mathcal{G}$
equipped with invariant inner product. Such a form can be considered as a
connection (gauge field) in a trivial vector bundle over $M$ and one can
modify (10) to get a functional that is invariant with respect to
infinitesimal gauge transformations $\delta A=d\gamma +[\gamma ,A]$ where $%
\gamma $ is a $\mathcal{G}$-valued function. This functional (Chern-Simons
action functional) has the form 
\begin{equation}
S(A)=\int_{M}(\frac{1}{2}A\wedge dA+\frac{1}{3}A\wedge A\wedge A)  \tag{11}
\end{equation}
where $A\wedge dA$ stands for $h_{ik}A^{i}\wedge (dA)^{k}$ and $A\wedge
A\wedge A$ stands for $f_{ijk}A^{i}\wedge A^{j}\wedge A^{k}$. (We denote by $%
A^{i}$ components of $A$ with respect to a basis in $\mathcal{G}$; $f_{ijk}$%
are structure constants of $\mathcal{G}$ and $h_{ik}$ is the metric tensor
of $\mathcal{G}$ in this basis.)

\bigskip

The Chern-Simons functional depends only on smooth structure of $M$
therefore one can hope it gives invariants of $M$. If $\Gamma $ is a
closed
curve in $M$ one can construct a gauge invariant expression $W_{\Gamma }(A)$
as trace of monodromy of the connection $A$ in some representation of the
group $G$ corresponding to the Lie algebra $\mathcal{G}$. Integral of $%
W_{\Gamma }(A)\exp (-kS(A))$ over infinite-dimensional space of all gauge
fields (or, better to say, over the space of gauge classes of gauge fields)
depends only on isotopy class of $\Gamma $ considered as a knot in $M$ and
on topology of $M$. There exist two ways to obtain well defined invariants
from this ill-defined integral: to use perturbation theory or to calculate
this integral precisely in terms of two-dimensional conformal theory. Direct
application of Faddeev-Popov procedure leads to very complicated expressions
that I have written down in 1987, but was not able to analyze rigorously.
Mathematical analysis of perturbation series was performed much later in %
\cite{as}, \cite{k1} on the base of diagram technique that can be obtained
from Batalin-Vilkovisky formalism (see Sec.7). Some remarks about Witten's
explicit solution are contained in Sec.8.

\bigskip

\textbf{3. Gauge theories in BRST-BFV formalism.}

\bigskip

Let us consider a $\mathbb{Z}_{2}$-graded vector space $\mathcal{E}$
equipped with an odd operator $\widehat{\Omega }$ obeying $\widehat{\Omega }%
^{2}=0$ (in mathematical terminology $\widehat{\Omega }$ is a differential,
in physics $\widehat{\Omega }$ is called BRST operator; BRST stands for
Becchi-Rouet-Stora-Tyutin). An operator $A:\mathcal{E}\rightarrow
\mathcal{E}$
commuting with $\widehat{\Omega }$ is called quantum observable; such an
operator descends to an operator $\widetilde{A}$ : $\widetilde{\mathcal{E}}%
\rightarrow \widetilde{\mathcal{E}}$ acting on homology $\widetilde{\mathcal{%
E}}=\func{Ker}\widehat{\Omega }/\func{Im}\widehat{\Omega }$. It is easy to
check that $\func{Tr}A=\func{Tr}\widetilde{A},$ where $\func{Tr}$ stands for
supertrace. (This fact is used in topology in the derivation of Lefschetz
fixed point formula.)

If $A$ can be represented as a (super)commutator of $\Omega $ with some
operator $B$ we can say that the observable $A$ is trivial: $\func{Tr}A=0$.
Quantum observables are called also BRST-closed operators, trivial
observables are BRST-exact. (Observables are related to the homology of the
space of linear operators on $\mathcal{E}$ where $\Omega $ acts by the
formula $A\rightarrow \lbrack \Omega ,A]$, where $[,]$ stands for
supercommutator.) If $A$ and $H$ both commute with $\widehat{\Omega }$ we
obtain that 
\begin{equation*}
\func{Tr}\widetilde{A}\exp (-\widetilde{H}\beta )=\func{Tr}A\exp (-H\beta ).
\end{equation*}%
This formula shows that at the level of expectation values of observables
the theory with Hamiltonian $H$ on $\mathcal{E}$ is equivalent to the theory
with Hamiltonian $\widetilde{H}$ on $\widetilde{\mathcal{E}}$. This
observation permits us to replace a theory with complicated space $%
\widetilde{\mathcal{E}}$ by a theory with simple space $\mathcal{E}$ at the
price of introducing additional degrees of freedom and BRST operator (see %
\cite{134} for more detail). One can say that BRST formalism is a version of
standard mathematical idea of resolution, when a complicated module is
replaced with a complex of simple modules.

Let us consider for example operators $T_{\alpha :}:E\rightarrow E,\alpha
=1,...,n,$ acting on space $E$ and generating a Lie algebra$\,\,\mathcal{G}$
(i.e. $[T_{\alpha },T_{\beta }]=f_{\alpha \beta }^{\gamma }T_{\gamma }$).
Let us denote by $\mathcal{E}$ the space of all $E$-valued functions
depending on anticommuting variables $c^{1},...,c^{n}$ (the space of
cochains of Lie algebra $\mathcal{G}$). The operator 
\begin{equation*}
\widehat{\Omega }=T_{\alpha }c^{\alpha }+\frac{1}{2}f_{\alpha \beta
}^{\gamma }c^{\alpha }c^{\beta }\frac{\partial }{\partial c^{\gamma }}
\end{equation*}%
obeys $\widehat{\Omega }^{2}=0$; corresponding homology $\widetilde{\mathcal{%
E}}=H(\mathcal{G},E)$ are called Lie algebra cohomology. The space $\mathcal{%
E}$, and, therefore, $\widetilde{\mathcal{E}}$ have natural $\mathbb{Z}$%
-grading with $\deg c^{\alpha }=1$. It is easy to see that $\widetilde{%
\mathcal{E}}^{0}=H^{0}(\mathcal{G},E)$ is the $\mathcal{G}$-invariant
subspace of $E$ (i.e. $\widetilde{\mathcal{E}}^{0}=\{x\in E:T_{\alpha }x=0\}$%
). Let $H$ be a $\mathcal{G}$-invariant hamiltonian on $E$. Then can
restrict it to $\widetilde{\mathcal{E}}^{0}$ (i.e we can introduce
constraints $T_{\alpha }x=0$.) We will assume that $\widetilde{\mathcal{E}}%
^{i}=H^{i}(\mathcal{G},E)=0$ for $i>0$ (the cohomology is concentrated in
degree $0$). Then the physics described by the Hamiltonian $H$ restricted to 
$\widetilde{\mathcal{E}}^{0}$ is equivalent to the physics of $H$ extended
to $\mathcal{E}$ if we are interested in expectation values of BRST-closed
operators. This is the easiest way to take into account constraints: instead
of restricting the space we enhance it including ghosts.

The quantum consideration above has a classical counterpart. In Hamiltonian
approach we should work with symplectic supermanifold $M$ (i.e. with a
supermanifold equipped with even close nondegenerate $2$-form). An analog of
a BRST-operator is a function $\Omega $ on $M$ obeying $\{\Omega ,\Omega
\}=0 $ where $\{,\}$ stands for Poisson bracket. Classical observables are
associated with homology of operator $\widehat{Q}:A\rightarrow \{A,\Omega \}$%
\bigskip\ acting on the space of functions on $M$.

The operator $\widehat{Q}$, obeying $\widehat{Q}^{2}=0$, can be considered
as an odd vector field on $M$. We say that such a vector field specifies a
structure of $Q$-manifold on $M$. Notice, that this structure is compatible
with symplectic structure (the Lie derivative of symplectic form with
respect to $Q$ vanishes).

Gauge theories in Hamiltonian formalism are systems with constraints. The
simplest way to study Hamiltonian constrained systems is to introduce ghosts
as we have explained. However, usually it is easier to work in Lagrangian
formalism. Lagrangian analogs of above constructions will be described in
the next section.

\bigskip

\textbf{4. Batalin-Vilkovisky formalism.}

Let us consider an $(n\mid n)$-dimensional supermanifold $M$ equipped with
an odd non-degenerate closed $2$-form $\omega =\omega _{ij}dz^{i}dz^{j}$.
We'll say that $M$ is an odd symplectic manifold (a $P$-manifold). In
appropriate local coordinates $\omega $ has the form $\omega =\sum
dx^{i}d\xi _{i}$, $i=1,...,n$; in other words $M$ can be pasted together
from $(n\mid n)$-dimensional superdomains by means of transformations
preserving $\sum dx^{i}d\xi _{i}.$

In the same way as on even symplectic manifold we can define Poisson bracket 
$\{f,g\}$ on a $P$-manifold. For every function $H$ on $P$-manifold $M$ we
define first order differential operator $\widehat{K}_{H}$ (a vector field $%
K_{H}$) by the formula $\widehat{K}_{H}(f)=\{f,H\}$.  It is easy to check
that $K_{H}$ preserves odd symplectic structure (i.e. Lie derivative $L_{K}$
of $\omega $ with respect to $K_{H}$ vanishes). Conversely, if a vector
field $K$ preserves $\omega ,$ it can be represented as $K_{H}$ at least
locally. If a $P$-manifold $M$ is equipped with a volume element we can
define an odd second order differential operator $\Delta $ on $M$ by the
formula $\Delta f=divK_{f}$. If the operator $\Delta $ obeys $\Delta ^{2}=0$
we say that $M$ is an $SP$-manifold. One can prove that an $SP$-manifold can
be pasted together from $(n\mid n)$-dimensional superdomains by means of
transformations preserving $\omega =\sum dx^{i}d\xi _{i}$ and volume
element; the operator $\Delta $ is equal to 2$\partial ^{2}/\partial
x^{i}\partial \xi _{i}$ in the coordinates $(x^{1},...,x^{n},\xi
_{1},...,\xi _{n}).$

\bigskip

Let us consider a function $A$ defined on a compact $SP$-manifold $M$and
obeying $\Delta A=0$. One can prove the following statement :

The expression 
\begin{equation}
\int_{L}Ad\nu  \tag{12}
\end{equation}%
where $L$ is a Lagrangian submanifold of $M$ does not change by continuous
variation of $L$; moreover, $L$ can be replaced by any other Lagrangian
submanifold $L^{^{\prime }}$belonging to the same homology class. (The
notion of Lagrangian submanifold of odd symplectic manifold can be defined
as in even case; a Lagrangian submanifold of $L$ of an $SP$-manifold can be
equipped naturally by a volume element $d\nu $.) In the case when $A=\Delta
B $ the integral (12) vanishes; this means that (12) determines a functional
on $Ker\Delta /Im\Delta $. If the function $A$ is represented in the form $%
A=\exp (\hbar ^{-1}S)$ the equation $\Delta A=0$ is equivalent to the
following equation for $S$%
\begin{equation}
\hbar \Delta S+\{S,S\}=0.  \tag{13}
\end{equation}%
This equation is known as quantum master equation; it plays an important
role in BV\ quantization procedure. Namely, in this procedure we can take as
a starting point classical action functional and construct a solution to
(13); the physical quantities are obtained as integrals of the form 
\begin{equation*}
\int_{L}e^{-\frac{1}{h}S}d\nu ;
\end{equation*}%
the choice of Lagrangian submanifold $L$ corresponds to the choice of gauge
condition. Of course, in the quantization problem we should consider
ill-defined infinite-dimensional integrals; statements about the integral
(12) are proved rigorously in finite-dimensional case, but don't have any
precise meaning in infinite-dimensional situation. Moreover, it is difficult
even to define the notion of infinite-dimensional $SP$-manifold and to
construct the operator $\Delta $. Nevertheless, one can use the framework of
perturbation theory to quantize gauge theories in BV formalism.

Let us emphasize that the notion of $P$-manifold and the equation 
\begin{equation}
\{S,S\}=0  \tag{14}
\end{equation}%
(classical master equation) make sense in the infinite-dimensional case. It
is natural to say that a solution of (14) specifies a classical mechanical
system in BV formalism. We will show that this viewpoint permits us to give
a very simple construction of topological quantum field theories.

First of all we should give a geometric interpretation of the solution to
classical master equation. Let us denote by $Q$ an odd vector field
corresponding to $S$. It follows from $\{S,S\}=0$ that $\{Q,Q\}=0$ (in other
words the first order differential operator $\widehat{Q}$ defined by the
formula $\widehat{Q}\Phi =\{\Phi ,S\}$ obeys $\widehat{Q}^{2}=0$). We'll say
that a supermanifold equipped with an odd vector field $Q$ obeying $%
\{Q,Q\}=0 $ is a $Q$-manifold. We see that a solution of classical master
equation on a $P$-manifold $M$ specifies a structure of a $Q$-manifold on $M$%
; these two structures are compatible (the odd symplectic structure is $Q$%
-invariant; i.e. the Lie derivative of odd symplectic form with respect to $%
Q $ vanishes). It is easy to check that, conversely, every $QP$-manifold
(i.e. $Q $-manifold equipped with $Q$-invariant odd symplectic structure)
can be obtained from a solution of classical master equation.

One can obtain many examples of $QP$-manifolds by means of simple geometric
constructions. We'll show how to construct topological quantum field
theories this way.

Notice, first of all that the space $\{\Sigma \rightarrow X\}$ of maps of a $%
Q$-manifold $\Sigma $ into a $Q$-manifold $X$ can be considered as a $Q$%
-manifold. Let us take as $\Sigma $ the manifold $\Pi TM$ where $M$ is a $d$%
-dimensional manifold (the symbol $\Pi $ stands for parity change; one
obtains the supermanifold $\Pi TM$ from the tangent bundle $TM$ reversing
parity of the fibers). The functions on $\Pi TM$ can be regarded as
differential forms on $M$; de Rham differential can be interpreted as an odd
vector field $Q$ on $\Pi TM$ obeying $\{Q,Q\}=0$. This means that $\Pi TM$
can be considered as a $Q$-manifold. The natural volume element on $\Pi TM$
is $Q$-invariant (i.e. $divQ=0$). Notice that the volume element on $\Pi TM$
is odd if $M$ is odd--dimensional and even if $M$ is even-dimensional.

To introduce a symplectic structure on the space of maps $\{\Sigma
\rightarrow X\}$ we need a volume element on $\Sigma $ and a symplectic
structure on $X$. Then the symplectic form on the space of maps can be
defined as an integral 
\begin{equation*}
\widetilde{\omega }(\delta _{1}f,\delta _{2}f)=\int_{\Sigma }\omega (\delta
_{1}f(\sigma ),\delta _{2}f(\sigma ))d\sigma
\end{equation*}
where $\omega $ stands for symplectic form on $X$ and $d\sigma $ for volume
element on $\Sigma $. To obtain an odd symplectic structure on $\{\Sigma
\rightarrow X\}$ we should assume that the parity of symplectic structure on 
$X$ is opposite to the parity of volume element on $\Sigma $.

Now we can say that $\{\Sigma \rightarrow X\}$ is a $QP$-manifold if $\Sigma 
$ is a $Q$-manifold equipped with even $Q$-invariant volume element and $%
\Sigma $ is a $QP$-manifold. If $\Sigma $ is a $Q$-manifold with odd $Q$%
-invariant volume element, then we can introduce a structure of a $QP$%
-manifold in $\{\Sigma \rightarrow X\}$ in the case when $X$ is a $Q$%
-manifold with $Q$-invariant even symplectic structure. The functional $S$
on $\{\Sigma \rightarrow X\}$ corresponding to the vector field $Q$ obeys $%
\{S,S\}=0$ and specifies a classical mechanical system, that can be called
BV sigma-model. In the case when $\Sigma =\Pi TM$corresponding sigma-model
can be considered as topological field theory\ (the action functional
depends only on smooth structure of the manifold $M$ and therefore
corresponding physical quantities should provide diffeomorphism invariants
of $M$). This general construction leads to many interesting TQFTs. In
particular, we can obtain (generalized) Chern-Simons theory in the following
way. Let us take as $X$ a linear $\Pi \mathcal{G}$ where $\mathcal{G}$ is a
Lie algebra equipped with invariant inner product. One can consider $\Pi 
\mathcal{G}$ as a $Q$-manifold equipped with $Q$-invariant symplectic
structure. Functions on $X$ can be interpreted as cochains of Lie algebra $%
\mathcal{G}$; the differencial acting on cochains can be considered as
vector field $Q$ on $X$. The symmetric inner product on $\mathcal{G}$
specifies an even symplectic structure on $X=\Pi \mathcal{G}$).

A map of $\Sigma =\Pi TM$ into $X=\Pi \mathcal{G}$ can be considered as $%
\mathcal{G}$-valued differential form $A$ on $M$. To obtain a structure of a 
$QP$-manifold on $\{\Pi TM\rightarrow \Pi \mathcal{G}\}$ we assume that $M$
is an odd-dimensional manifold. Above arguments lead to the following action
functional: 
\begin{equation}
S(A)=\int_{M}(\frac{1}{2}A\wedge dA+\frac{1}{3}A\wedge A\wedge A)  \tag{15}
\end{equation}%
In the case $\dim M=3$ we obtain a BV version of Chern-Simons action
functional.

Let us consider the space $\{\Pi TM\rightarrow X\}$ where $M$ is an
odd-dimensional manifold and $X$ is an even symplectic manifold equipped
with trivial $Q-$structure ($Q=0$). We will analyze in detail the case when $%
X$ is a symplectic vector space with symplectic form having constant
coefficients $\omega _{\alpha \beta }$. In this case maps $\Pi TM\rightarrow
X$ can be identified with vector valued forms $A^{\alpha }$ on $M$ ($\alpha
=1,...,\dim X$) and the action functional is quadratic.%
\begin{equation}
S=\int_{M}\omega _{\alpha \beta }A^{\alpha }\wedge A^{\beta }.  \tag{16}
\end{equation}

We obtain a BV-version of action functional (10); corresponding partition
function is related to Ray-Singer torsion.

For every manifold $Y$ we can construct a structure of a $Q$-manifold on $%
\Pi TY$ and a structure of a $P$-manifold on $\Pi T^{\ast }Y$. If $Y$ is an
even symplectic manifold we can identify $\Pi TY$ and $\Pi T^{\ast }Y$ and
obtain a structure of a $QP$-manifold on $\Pi T^{\ast }Y$; corresponding
solution to classical master equation can be written in the form $s=\omega
^{\alpha \beta }(y)\eta _{\alpha }\eta _{\beta }$, where $y\in Y$, $\eta
_{\alpha }$ are odd coordinates in the fibres of $\Pi T^{\ast }Y$ and $%
\omega ^{\alpha \beta }$ stands for bivector that is inverse to symplectic
form $\omega _{\alpha \beta }$. It is easy to verify that this construction
specifies a structure of a $QP$-manifold on $\Pi T^{\ast }Y$ in more general
case when $Y$ is a Poisson manifold ($s=\omega ^{\alpha \beta }(y)\eta
_{\alpha }\eta _{\beta }$ obeys classical master equation iff $\omega
^{\alpha \beta }$ determines Poisson structure on $Y$).We can consider now
topological BV sigma-model on the space of maps $\{\Pi TM\rightarrow \Pi
T^{\ast }Y\}$ where $M$ is a two-dimensional (or, more generally,
even-dimensional) manifold and $Y$ is a Poisson manifold. In the case when $%
M $ is a disk this sigma-model was used by Kontsevich in his famous work
about formal quantization of Poisson manifolds \cite{k2}, see also
\cite{CF}. Taking a symplectic manifold equipped with an almost
complex structure as $Y$ we can single out a Lagrangian submanifold $L$ of $%
\{\Pi TM\rightarrow \Pi T^{\ast }Y\}$ in such a way that restriction
functional to $L$ leads so called $A$-model \cite{aksz}.

\bigskip

\textbf{5.Perturbation theory in BV-formalism.}

\bigskip

Let us begin with finite-dimensional case. We consider a functional $S$
defined on linear $SP$-manifold $\mathcal{E}$. (Every manifold of this kind
is isomorphic to $\mathbb{R}^{n\mid n}$ equipped with standard odd
symplectic form and volume element.) We represent $S$ as $S_{0}+V$ where $%
S_{0}$ consists of quadratic terms in Taylor series at stationary point $%
x_{0}$ of $S$; without loss of generality we can take $x_{0}=0$. We assume
that $S$ obeys both quantum and classical master equations (i.e. $\{S,S\}=0$%
, $\Delta S=0$). Then $S_{0}$ and $V$ also have this property: $%
\{S_{0},S_{0}\}=0$, $\Delta S_{0}=0$, $\{V,V\}=0$, $\Delta V=0$; therefore $%
\{S_{0},V\}=0$. Let us consider a linear Lagrangian subspace $L\subset 
\mathcal{E}$; we suppose that $S_{0}$ is nondegenerate on $L$ and the
integral 
\begin{equation*}
Z(\lambda )=\int_{L}e^{-(S_{0}+\lambda V)}d\upsilon ,
\end{equation*}%
representing the partition function of action functional $S_{0}+\lambda V$,
converges. We can apply standard methods to get a perturbative expansion of $%
Z(\lambda )$ $/Z(0)$ with respect to $\lambda $. The series we obtain does
not depend on the choice of $L$; therefore one can describe the answer in a
form where $L$ is not involved. We'll prove that one can use the standard
Feynman diagram technique where vertices are governed by $V$ and the
propagator $p$ is a bivector on $\mathcal{E}$ that obeys 
\begin{equation}
q_{\gamma }^{\alpha }p^{\gamma \beta }-p^{\alpha \gamma }q_{\gamma }^{\beta
}=\omega ^{\alpha \beta }  \tag{17}
\end{equation}%
Here $\omega ^{\alpha \beta }$ is the bivector that is inverse to the matrix 
$\omega _{\alpha \beta }$ of odd symplectic form, $q_{\beta }^{\alpha
}=\omega ^{\alpha \gamma }s_{\gamma \beta }$ where $s_{\alpha \beta }$
stands for the matrix of quadratic form $S_{0}$. In more invariant form we
can say that quadratic form $S_{0}$ generates a vector field $Q$ on $%
\mathcal{E}$. The coordinates of $Q$ are linear functions on $\mathcal{E}$,
therefore we can construct a linear operator $q$ acting on $\mathcal{E}$; it
follows from $\{S_{0},S_{0}\}=0$ that $q^{2}=0$. This means that we can
regard $q$ as a differential; it follows from our assumptions that
corresponding homology is trivial ($\func{Ker}q=\func{Im}q$). The operator $%
q $ generates a differential in the space of bivectors; we'll use the same
symbol for it. The condition (17) means that $qp=\omega $ where $\omega $
stands for bivector, that is inverse to symplectic form. Notice that for two
bivectors $p^{\prime }$ and $p$ that obey (17) we have $q(p^{\prime
}-p)=0$.
It follows from acyclicity of $q$ that there exists an odd bivector $u$
satisfying $p^{\prime }-p=qu$. One can use this remark to show that diagrams
constructed by means of propagator $p^{\prime }$ coincide with diagrams with
propagator $p$.

\bigskip

The above prescription can be justified in the following way. One can
introduce such a coordinate system $x^{1},...,x^{n},\xi _{1},...,\xi _{n}$
on $\mathcal{E}$ that $S_{0}$ depends only on $\xi _{1},...,\xi _{n};$ the
parity of $\xi _{i}$ is opposite to the parity of $x^{i},$ the odd
symplectic form is equal to $\sum dx^{i}d\xi _{i}$ and the volume element is
standard (see \cite{4}). Then the partition function can be written as
integral of $\exp (-(S_{0}+\lambda V))$ over Lagrangian submanifold $x=X$,
where $X^{1},...,X^{n}$ is a fixed vector. This integral can be converted
into an integral over $\mathcal{E}$:%
\begin{equation}
Z=\limfunc{const}\int e^{-(S_{0}+\sigma )-\lambda V}d\xi dx  \tag{18}
\end{equation}%
(we included $\delta (x-X)$ into the integrand, multiplied it by $\exp
(-\sigma (X))$ and integrated over $X$. Here $\sigma (X)$stands for
nondegenerate quadratic form; we require convergence of the integral of $%
\exp (-\sigma (X))$ over $X^{1},...,X^{n}$).

It follows from (18) that $Z(\lambda )/Z(0)$ can be represented by means
of
Feynman diagrams where propagator is inverse to $S_{0}+\sigma $. It is easy
to check that this recipe coincides with the above prescription for specific
choice of bivector $p$ obeying (17). This means that we can use any
bivector
satisfying (17). (We mentioned already that diagrams don't depend on the
choice of propagator as long it obeys (17).)

We can apply the diagram technique developed in finite-dimensional case to
infinite-dimensional situation.

\bigskip

Let us consider for example the BV-formalism of Chern-Simons action
functional (i.e. functional (15) for $\dim M=3$). Then we obtain precisely
diagrams constructed by M. Kontsevich \cite{k1}. It is necessary to
emphasize,
however, that our considerations in infinite-dimensional  case are
heuristic.
To obtain rigorous results one should analyze the convergence of integrals
representing the diagrams,etc. (see \cite{k1}).

\bigskip

\bigskip \textbf{6.Observables .}

Let us consider an $SP$-manifold $M$ and a quantum system corresponding to a
solution $S$ of quantum master equation (13). We say that a function $A$ on $%
M$ is a quantum observable if it satisfies the equation

\begin{equation}
\hbar \Delta A+2\{A,S\}=0  \tag{19}
\end{equation}

It is important to notice that $A$ is not necessarily an even function. The
expression%
\begin{equation*}
\int_{L}Ae^{S/\hbar }d\nu
\end{equation*}
where $L$ is Lagrangian submanifold of $L$, has the meaning of the
expectation value of $A$. This expression depends only on homology class of $%
L$. For even $A$ this fact immediately follows from the remark that the
equation (17)\ holds that iff $S+\varepsilon A$ where $\varepsilon $ is an
infinitesimal parameter obeys quantum master equation; analogous statement
is true for odd $A$. If a quantum observable can be represented in the form $%
A=\hbar \Delta B+2\{B,S\}$ then its expectation value vanishes; we say that
such an observable is trivial.

The above remarks show that observables can be studied in the framework of
families of quantum systems; expectation values of observables govern the
variation of partition function by the infinitesimal change of paramaters.

One can prove the following statements:

a) If $A$ and $B$ are quantum observables, then $\{A,B\}$ is also a quantum
observable. (In other words, quantum observables constitute a Lie
superalgebra.)

b) Let us suppose that quantum observables $T_{\alpha }$ span a Lie
(super)algebra $\mathcal{G}$ (i.e. $\{T_{\alpha },T_{\beta }\}=f_{\alpha
\beta }^{\gamma }T_{\gamma }$) and that the antisymmetric tensor $c^{\alpha
_{1}...\alpha _{k}}$ \ represents a homology class of $\ \mathcal{G}$. Then$%
\ c^{\alpha _{1}...\alpha _{k}}T_{\alpha _{1}}...T_{\alpha _{k}}$ is also a
quantum observable. If the tensor $c^{\alpha _{1}...\alpha _{k}}$ belongs to
the trivial homology class, the corresponding observable is also trivial.

One can derive these statements by means of straightforward calculations
based on the definition of homology, on the relation 
\begin{equation*}
\{S,KL\}=\{S,K\}\cdot
L+(-1)^{\varepsilon (K)}\{S,L\}\cdot K
\end{equation*}
and on the formula%
\begin{equation*}
\Delta (KL)=\Delta K\cdot L+(-1)^{\varepsilon (K)}K\cdot \Delta
L+(-1)^{\varepsilon (K)}\{K,L\}.
\end{equation*}

\bigskip Taking $\hbar =0$ in the definition of quantum observable we obtain
a definition of classical observable. We can also consider functionals that
verify the condition (19) for all $\hbar $. One can say that these
functionals are quantum and classical observables at the same time; we will
omit the adjective talking about observables of this kind. These observables
are related to infinitesimal variations of action functionals obeying
quantum and classical master equations simultaneously: $\Delta S=0$, $%
\{S,S\}=0$. As we emphasized, in infinite-dimensional case the operator $%
\Delta $ is ill-defined, therefore it is difficult to work with quantum
master equation. It is much easier to make sense of equation $\Delta S=0$;
it can be written in the form $divQ=0$ and means that the odd vector field $%
Q $ corresponding to the functional $S$ is ''volume preserving.''

In Sec. 7 we'll discuss how to obtain topological invariants in the
framework of perturbation theory taking as a starting point the BV version
of
Chern-Simons action functional (15) .Notice that it is possible to obtain
perturbative Chern-Simons invariants also from quadratic action functional
(16) considering non-trivial quantum observables. Lie algebra $\mathcal{H}$
of polynomial Hamiltonian vector fields on X can be considered as an algebra
of symmetries of the functional (16). This means that we can associate a
quantum observable with every homology class of this Lie algebra.
Corresponding expectation values are topological invariants of $M$.
Kontsevich constructed a graph complex having homology closely related to
the homology of Lie algebra $\mathcal{H}$ (see \cite{k1}). It follows from
this result that one can associate topological invariants of $M$ with
homology of graph $\func{complex}$. It is shown in \cite{999} that
invariants
obtained this way coincide with invariants derived in \cite{k1} from the
analysis of perturbative Chern-Simons theory. One can say that these
observations give physical explanation of some results of \cite{k1}.

\bigskip

\bigskip \textbf{7.Families of action functionals.}

\bigskip

Let us consider a smooth family $S_{\lambda }$ of functionals defined on
manifold $M$ and labeled by points $\lambda \in \Lambda $. We assume that
these functionals obey quantum and classical master equations for every $%
\lambda \in \Lambda $: 
\begin{equation*}
\Delta S_{\lambda }=0,\{S_{\lambda },S_{\lambda }\}=0.
\end{equation*}

Let $V$ stand for a vector on $\Lambda $. Then the variation of $S$ in
the
direction $V$ is governed by \bigskip oVbservable $T(V):$%
\begin{equation*}
\widehat{V}S=T(V),\Delta T(V)=0,\{T(V),S\}=0.
\end{equation*}%
We will assume that there exists functional $B(V)$ on $M$ \ obeying 
\begin{equation}
T(V)=\{B(V),S\},\Delta B(V)=0.  \tag{20}
\end{equation}%
Then the observable $T(V)$ is trivial, because 
\begin{equation*}
\widehat{V}e^{S}=T(V)e^{S}=\{B(V),e^{S}\}=\Delta (B(V)e^{S})
\end{equation*}

This means that the partition function corresponding to the action
functional $S_{\lambda }$ \ does not depend on $\lambda $. It seems that the
consideration of the family $S_{\lambda }$ is superfluous and we can
restrict ourselves to one of the members of this family. We'll see that this
is not the case. The functional $T(V)$ is defined for every vector $V$ on
(super)manifold $\Lambda $. For definiteness we consider only even vectors;
then $T(V)$ is an even functional and $B(V)$ is an odd functional. We will
use the notation $B(V)$ also in the case when $V$ is a vector field; then $%
B(V)$ is a function depending on $x\in M$ and $\lambda \in \Lambda $. It is
easy to check that 
\begin{equation*}
\widehat{V}_{1}\widehat{V}_{2}e^{S}=\widehat{V}_{1}\{B(V_{2}),e^{S}\}=\{%
\widehat{V}_{1}B(V_{2}),e^{S}\}+\{B(V_{2}),\{B(V_{1}),e^{S}\}\}.
\end{equation*}%
Comparing this formula with 
\begin{equation*}
\lbrack \widehat{V}_{1},\widehat{V}_{2}]e^{S}=\{B([V_{1},V_{2}]),e^{S}\}
\end{equation*}%
we obtain that 
\begin{equation*}
B([V_{1},V_{2}])=\widehat{V}_{1}B(V_{2})-\widehat{V}_{2}B(V_{1})-%
\{B(V_{1}),B(V_{2})\}+\xi (V_{1},V_{2})
\end{equation*}%
where $\Delta \xi (V_{1},V_{2})=0,\{\xi (V_{1},V_{2}),S\}=0.$

We see that $\xi (V_{1},V_{2})$ is an observable.

We consider $\xi (V_{1},V_{2})$ as a function on $M$ and two-form on $%
\Lambda $; one can check that this two-form is closed. In most interesting
cases this form can be represented as a differential of one-form $\eta $
obeying $\Delta \eta =0,\{\eta (V)),S\}=0$. If this exactness condition is
satisfied we can replace $B(V)$ with $B(V)-\eta (V)$\ preserving relations
(20); with this new definition of $B(V)$ the form $\xi (V_{1},V_{2})$
vanishes, i.e. 
\begin{equation}
B([V_{1},V_{2}])=\widehat{V}_{1}B(V_{2})-\widehat{V}_{2}B(V_{1})-%
\{B(V_{1}),B(V_{2})\}.  \tag{21}
\end{equation}%
We'll assume that (21) is satisfied. Then one can prove that the $n$-form
$%
\Delta \omega _{n}=\Delta (B(V_{1})...B(V_{n})e^{S})$ is a differential of $%
(n-1)$-form $\omega _{n-1}=B(V_{1})...B(V_{n-1})e^{S}$ on $\Lambda $. This
means that for every $n$-cycle $\Gamma $ on $\Lambda $ we have 
\begin{equation*}
\Delta \int_{\Gamma }B(V_{1})...B(V_{n})e^{S}=0.
\end{equation*}%
Integrating $\int_{\Gamma }B(V_{1})...B(V_{n})e^{S}$ over a Lagrangian
submanifold $L\subset M$ we obtain a number that depends only on homology
classes of $\Gamma $ and $L.$

The above statement can be generalized to the case when $\xi (V_{1},V_{2})$
does not vanish. In this case it is convenient to consider an inhomogeneous
form $\omega =\sum \omega _{n}$. One can prove, that 
\begin{equation*}
\Delta \omega =(d+\xi )\omega .
\end{equation*}

\bigskip

Notice that instead of family $S(\lambda )$ of equivalent action functionals
we can consider a functional $S$ obeying $\Delta S=0,\{S,S\}=0,$ and a
family of Lagrangian submanifolds $L_{\lambda }$ labeled by $\lambda \in
\Lambda .$

\bigskip

Let us denote by $\mathcal{L}$ the infinite-dimensional manifold of all
Lagrangian submanifolds of $M$. (If $L\subset M$ is a Lagrangian
submanifold
we can identify a neighborhood of $L$ in $M$ with $\Pi T^{\ast }L$. Using
this identification we can construct for every odd function $\Psi $ on $L$ a
Lagrangian submanifold by means of the formula \ $\xi _{i}=\frac{\partial
\Psi }{\partial x^{i}}$ where \ $x^{i}$ are coordinates on \ $L$, $\xi _{i}$
are coordinates on the fibers. This construction gives a parametrization of
a subset of $\mathcal{L}$ in terms of functions on $L$; we see that $%
\mathcal{L}$ can be considered as infinite-dimensional manifold. Lie algebra
of even vector fields preserving odd symplectic structure and volume element
on $M$ acts on $\mathcal{L}$ in natural way; it is easy to check that this
action is Lie algebra of odd functionals $B$ on $M$ obeying $\Delta B=0$.
This means that to every $B$ obeying $\Delta B=0$ and every $L\in \mathcal{L}
$ corresponds a vector $V\in T_{L}(\mathcal{L})$ (a tangent vector at the
point $L\in \mathcal{L}$) and that this map is surjective. The inverse map
is multivalued, but one can fix one-valued smooth branch $B(V);$we'll use
the same notation when $V$ is a vector field on $\mathcal{L}$ . We'll assume
that 
\begin{equation*}
B([V_{1},V_{2}])=\widehat{V}_{1}B(V_{2})-\widehat{V}_{2}B(V_{1})-%
\{B(V_{1}),B(V_{2})\}.
\end{equation*}
as in the case of family of action functionals considered above.

Let us consider an $n$-form on $\mathcal{L}$ defined by the formula%
\begin{equation*}
\omega _{n}=\int_{L}B(V_{1})...B(V_{n})e^{S}d\nu
\end{equation*}
One can prove that this form is closed. The proof is based on the relation 
\begin{equation*}
\widehat{V}\int_{L}\varphi d\nu =\int_{L}\{\varphi ,B(V)\}d\nu
\end{equation*}

(We consider infinitesimal transformation of $M$ preserving volume element
and odd symplectic structure. To calculate the variation of $\int_{L}\varphi
d\nu $\ \ by the variation of $L$ we use the fact that instead of changing $%
L $ we can change the integrand.)

Notice that we can modify the definition of the forms $\omega _{n}$
including an observable \ $A$ \ into the integrand. The forms remain closed
after such a modification (this follows immediately from the remark that
observables are related to infinitesimal variations of action functional).

One can consider $\mathcal{L}$ or the space of equivalent action functionals
as a kind of moduli space for the problem at hand. We obtained under certain
conditions closed forms on this space. Integrating these forms over cycles
in moduli space we obtain numbers that generalize expectation values of
observables (these expectation values correspond to $0$-forms).

Let us apply the above consideration to the case of topological theories. In
this case every metric on the worldsheet determines one of equivalent action
functionals (in Witten's approach) or a gauge condition (a Lagrangian
submanifold in BV approach).Under certain conditions closed forms on the
space of metrics are equivariant with respect to diffeomorphisms of
worldsheet $M$ and therefore descend to the quotient space $\mathcal{M}$%
=\{metrics\}/\{diffeomorphisms\}. The space of metrics is contractible hence
the quotient space is closely related to the classifying space BDiff($M$) of
diffeomorphism group. In the case of multidimensional version of BV
Chern-Simons functional the differential forms constructed above are related
to forms on BDiff($M$) considered in [8]. (Notice that Kontsevich modifies
the space BDiff ($M$) to get rid of quantum anomalies connected with
zero-dimensional homology of $M$.) In the case of two-dimensional
topological theory the numbers obtained by means of integration of
differential forms on moduli space over cycles coincide with string
amplitudes. Notice, that in two-dimensional case the moduli space of metrics 
$\mathcal{M}$ is homotopy equivalent to\ the moduli space of conformal
structures on \ the worldsheet (=moduli space of complex curves with given
topology). The appearance of so called Deligne-Mumford compactification of
the moduli space of complex curves is related to the fact that one can
obtain reasonable gauge conditions allowing metrics with some mild
singularities.

\bigskip

\textbf{8. Chern-Simons theory and topological sigma-model}.

\bigskip

Chern-Simons theory is closely related to so called $G/G$\ model. \ This
two-dimensional topological model can be considered as gauged WZNW model and
can be solved either by conformal field theory methods or directly. To
establish the relation between of Chern-Simons theory \ and $G/G$ model one
can use WZNW model as an intermediate step (see \cite{W4}); there exists
also
more direct way found in \cite{bl}.

All these approaches are based \ on the remark that $1$-form $A$ satisfying
the equations of motion corresponding to Chern-Simons action functional
(11)
can be considered as a flat connection on trivial vector bundle.  In
topologically trivial situation all flat connections are gauge equivalent;
this means that every flat connection can be represented in the form

\begin{equation*}
A=g^{-1}(x)dg(x) \tag{21}
\end{equation*}%
where $g(x)$ is a function \ taking values in the group \ $G$ . (We suppose
that $\mathcal{G}$ is a Lie algebra of the group $G$.) \ \ \ \ \ \ \ Another
method uses BV-formalism; this approach can be applied also to
multidimensional generalization (15) of  \ \ Chern-Simons \ action
functional.\ It is based on the remark that equation of motion corresponding to
the\ action functional (15) can be easily solved. These equations have the
form 
\begin{equation*}
dA+A\wedge A=0 \tag{22}
\end{equation*}%
where $A$ is $\Pi \mathcal{G}$-valued function on $\Pi TM\ $($\mathcal{G}$%
-valued inhomogeneous form on $M$). 

To every $G$-valued function $g(x)$\ on $\Pi TM$ we can assign a $\Pi 
\mathcal{G}$-valued function on $\Pi TM$ by formula $A=g^{-1}(x)(Qg)(x)$ or,
more precisely,

\begin{equation*}
A=(g^{-1}(x))_{\ast }(Qg)(x).\tag{23}
\end{equation*}

Recall that $\Pi TM$ is a $Q$-manifold. The vector field $Q$ on $\Pi TM$
determines a vector field on the space of maps $\{\Pi TM\rightarrow G\}$
that is denoted by the same letter. The symbol $(g^{-1}(x))_{\ast }$ stands
for the map of tangent space $g(x)$ of the space $\{\Pi TM\rightarrow G\}$
at the point into tangent space at the point $g(x)=1$ that is induced by
left multiplication: $h(x)\rightarrow g^{-1}(x)h(x)$.

It is easy to check that (23) satisfies Eqn (22) and that in topologically
trivial situation all solutions to (22) can be obtained this way.
\bigskip

\textbf{Acknowledgements}

I am deeply indebted to M. Kontsevich for useful comments.

\end{document}